\documentclass{siamonline190516}

\usepackage{amsmath,amssymb,graphicx,color}
\usepackage{tikz}
\usepackage{dsfont}
\usepackage{mathtools}
\usetikzlibrary{shapes.geometric, arrows}
\usepackage{algorithmic}
\usepackage[english]{babel}
\usepackage[utf8]{inputenc}
\usepackage{mathrsfs}  
\usepackage{afterpage}
\usepackage{mathtools}
\usepackage{amsfonts}
\usepackage[colorinlistoftodos]{todonotes}
\usepackage{enumerate}
\usepackage[round]{natbib}
\usepackage{hyperref}
\usepackage{comment}

\newcommand{\excise}[1]{}

\newcommand\RR{\mathbb{R}}

\newcommand\MM{\mathcal{M}}

\newcommand{\eps}{\varepsilon}

\usepackage[mathlines]{lineno}

\begin{document}
	
	\title{\mbox{}\\[-11ex]Exponential-wrapped distributions on symmetric spaces}
	\author{\vspace{1em}Emmanuel Chevallier$^{1}$, Didong Li$^{2}$, Yulong Lu$^{3}$ and David Dunson$^{4,5}$  \\ 
		{\em Aix Marseille Univ, CNRS, Centrale Marseille, Institut Fresnel$^1$}\\
		{\em Department of Biostatistics, University of North Carolina at Chapel Hill$^{2}$}\\
		{\em Department of Mathematics and Statistics, University of Massachusetts Amherst$^{3}$}\\
		{\em Department of Statistical Sciences$^{4}$ and Mathematics$^{5}$, Duke University}
	}
	
	\maketitle
	
	\begin{abstract}
		In many applications, the curvature of the space supporting the data makes the statistical modelling challenging. In this paper we discuss the construction and use of probability distributions wrapped around manifolds using exponential maps. These distributions have already been used on specific manifolds. We describe their construction in the unifying framework of affine locally symmetric spaces. Affine locally symmetric spaces are a broad class of manifolds containing many manifolds encountered in data sciences. We show that on these spaces, exponential-wrapped distributions enjoy interesting properties for practical use. We provide the generic expression of the Jacobian appearing in these distributions and compute it on two particular examples: Grassmannians and pseudo-hyperboloids. We illustrate the interest of such distributions in a classification experiment on simulated data.

	\end{abstract}
	
	\section{Introduction}
	
	Density estimation on manifolds has been the subject of theoretical studies for several decades \citep{hall1987kernel,hendriks1990nonparametric,kim1998deconvolution,pelletier2005kernel,huckemann2010mobius}. More recently, probability densities on manifolds have also become tools of major interest in applied data science, from classification of video data on Grassmannian manifolds \citep{turaga2011statistical,slama2015accurate}, to modeling of hierachical structures on hyperbolic spaces \citep{ding2020deep,mathieu2019continuous}. An important difficulty is to define statistical models adapted to practical use for broad classes of manifolds; much of the focus has been on developing methods for specific manifolds, such as the sphere \citep{fisher1953dispertion,hauberg2018gaussian,kato2020shogo}.
	On a Riemannian manifold, a seemingly simple candidate family includes distributions whose densities with respect to the Riemannian measure are the normalized indicator functions
	\begin{equation*}
		\label{uniformKernel}
		\varphi_{p,r}= \frac{1}{v(\mathcal{B}(p,r))} \mathds{1}_{\mathcal{B}(p,r)}, 
	\end{equation*}
	where $\mathcal{B}(p,r)$ is the ball centered at $p$ of radius $r$ and $v$ is the Riemannian volume. However, computing the normalization constant is typically non-trivial, as there are no closed form expressions for the volume of balls.

	In this article, we focus on statistical models defined by pushing probability densities supported on tangent spaces to the manifold, using an exponential map. Since there exist various ways to push a density from a tangent space to the manifold; we refer to such densities as ``exponential-wrapped densities''. Exponential-wrapped densities have been studied and used in many applications, see for instance \cite{pelletier2005kernel,falorsi2019reparametrizing,mathieu2019continuous,ding2020deep,mallasto2019probabilistic,mallasto2019wrapped,kurtek2012statistical,turaga2011statistical,srivastava2005statistical,slama2014grassmannian,slama2015accurate,jona2012spatial,chevallier2015probability,chevallier2016kernel}.
	Most of these papers focus on individual manifolds, where the exponential-wrapped densities enjoy interesting properties. Our over-arching contribution is to develop a unified framework, and corresponding theory and methodology, for exponential-wrapped modeling on affine locally symmetric spaces (ALSS). ALSS encompass most manifolds used in data science, including (pseudo-)Riemannian symmetric spaces and arbitrary Lie groups. For reasons mentioned later in the introduction, ALSS are likely to form the most general setting on which exponential-wrapped densities remain tractable.
	
	Defining an exponential-wrapped density requires the existence of an exponential map. The exponential map is commonly defined for Riemannian manifolds using geodesics, or for Lie groups using one-parameter subgroups. However, both exponentials can be seen as exponential maps of an underlying affine connection. In this article, symmetric spaces refer to affine symmetric spaces in general, and not to Riemannian symmetric spaces. 
	
	Manifolds with affine connections are to Riemannian manifolds what affine spaces are to Euclidean vector spaces: they have a notion of straight lines but no distance. It is interesting to note that many statistical models on $\mathbb{R}^n$ do not depend on the Euclidean structure. For instance, defining a Gaussian distribution relies only on the affine structure and not on the distance. Similarly, exponential-wrapped models on Riemannian manifolds usually depend only on the affine connection associated with the metric. The main difference between the two settings is that the affine structure does not provide a notion of isotropic distributions. 
	
\begin{figure}\label{Fig:volumeChange}
		\begin{center}
			\includegraphics[scale=0.5]{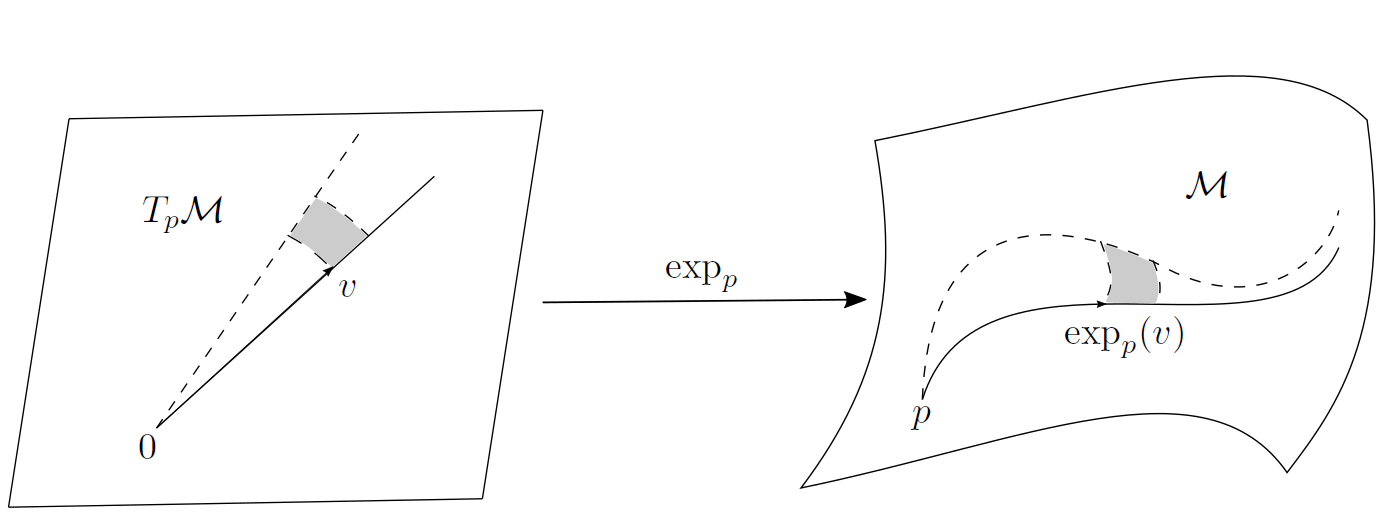}
			\caption{The infinitesimal volume change between the grey areas depends on how the neighboring geodesics are deviating or getting closer. It is given by the absolute value of the Jacobian determinant of the exponential map, and is determined by the curvature of the connection $\nabla$ along the geodesic $\exp_p(tv)$.}
		\end{center}
	\end{figure}

	In order to obtain tractable exponential-wrapped densities, it is important that the exponential map, its inverse, and its Jacobian determinant, see Figure \ref{Fig:volumeChange}, admit simple expressions. ALSS provide a broad class in which this is possible.  First, exponential maps and their inverses on injectivity domains can be computed at a reasonable cost: they can be identified to a Lie group exponential. Second, we provide explicit expressions of the Jacobian determinants for arbitrary symmetric spaces. The differential of the exponential map is governed by a matrix second order differential equation: the equation of Jacobi fields. On locally symmetric spaces this equation has constant matrix coefficients, which enables the computation of the Jacobian determinant. 
	
	Outside of ALSS, we expect this to happen on only a few specific manifolds. We are currently aware of only two examples of non locally symmetric manifolds appearing in data science where the exponential map, its inverse and Jacobian determinant can also be computed easily: Gaussian distributions endowed with the Wasserstein Riemannian metric \citep{chevallier2017kernel} and Kendall shape spaces  \citep{nava2019geodesic}. 

	In section \ref{sec:wrappedDistributions}, we describe exponential-wrapped distributions on ALSS. This setting encompasses and generalizes most previously considered settings, while preserving all the advantages of wrapped distributions.
	In section \ref{sec:SymmetricSpaces} we give the formal definitions of affine locally symmetric spaces and homogeneous symmetric spaces, and set some notations.
	In section \ref{sec:TheJacobian} we provide the general expression of the Jacobian appearing in exponential-wrapped densities on ALSS, and compute it on two original examples: Grassmannian manifolds and pseudo-hyperboloids.
	In section \ref{sec:A classification experiment using exponential wrapped distributions}, we present a classification experiment based on exponential-wrapped distributions. The experiment shows the interest of using multiple tangent spaces to model data. Section \ref{sec:Discussion} concludes the paper.

	\section{Exponential-wrapped densities}\label{sec:wrappedDistributions}
	
	Exponential-wrapped densities are traditionally used to define distribution on the circle $S_1$, see for instance \cite{mardia1972statistics} page 53. The density on the circle is obtained by taking a density on $\mathbb{R}$ and by wrapping it around a circle. Formally, if $f$ is a density on $\mathbb{R}$, the wrapped density can be defined as
	$$ f_{2\pi}(\theta) = \sum_{k=-\infty}^{\infty} f(\theta + k 2 \pi).$$
	
	Wrapped densities on circle can sometimes be written in closed form, it is the case for instance when $f$ is a Cauchy distribution. When the circle $S_1$ is viewed as a Riemannian manifold and $\mathbb{R}$ as a tangent space, the map $x\mapsto x \mod{2\pi}$ can be interpreted as a Riemannian exponential map. This point of view enables extension to more general manifolds endowed with an exponential map. In the vocabulary of measure theory, the exponential-wrapped probability is the pushforward of the probability in the tangent space by the exponential map. When the dimension of the space is greater than one, wrapping a density from a tangent space around the manifold usually requires taking into account a volume distortion. Indeed, the exponential map is generally not an area preserving map between the tangent space with a Lebesgue measure and the reference measure on the manifold. In this paper, we focus on the cases where the probability distributions in the tangent spaces are contained in injectivity domains of the exponential maps. This is a restrictive assumption on manifolds such as spheres, where the injectivity domains are disks. However, as we will see in section \ref{subsec:Isotropic exponential-wrapped normal distributions}, it holds, at least approximately, for most exponential-wrapped distributions used in practice. In this context, the difficulty does not lie in the computation of an infinite series as for most standard wrapped densities on circles, but in the computation of the volume distortion.
	
	Start by giving a precise definition of exponential-wrapped distributions. Let $\mathcal{M}$ be a manifold with a reference measure $vol$, and an exponential map $\exp_p:T_p\mathcal{M}\rightarrow \mathcal{M}$ at $p$, a point in $\mathcal{M}$. Given $\lambda$, a probability distribution on $T_p\mathcal{M}$, the corresponding exponential-wrapped distribution is defined as the push-forward of $\lambda$ by the exponential:
	\begin{equation}
		\label{eq:wrappedDistributions}  
		\Lambda=\exp_{p*}\lambda,
	\end{equation}
	
	where the $*$ refers to the push-forward by $\exp_p$: $\Lambda(A) = \lambda(\exp_p^{-1}(A)) $. In the rest of the paper, we assume that $\lambda$ is supported on a domain $U\subset T_p\mathcal{M}$ on which $\exp_p$ is injective, and that it has a density $h$ with respect to a Lebesgue measure $\nu_p$ of $T_p\mathcal{M}$. Under these assumptions, the density $f$ of $\Lambda$ can be expressed from $h$ and a volume change term. When $q=\exp_u$, we have
	
	\begin{equation}
		\label{eq:wrappedDensity}
		f(q) = \frac{\mathrm{d}\Lambda}{\mathrm{d}vol }(q) = \frac{\mathrm{d}\exp_{p*}(\nu_p)}{\mathrm{d}vol} \frac{\mathrm{d}\Lambda}{\mathrm{d}\exp_{p*}(\nu_p)}(q)  = \frac{\mathrm{d}\exp_{p*}(\nu_p)}{\mathrm{d}vol}(q)h(u),
	\end{equation}
	
	and when $q\notin \exp_p(U)$, $f(q)=0$.
	The volume change term is determined by the Jacobian determinant of the differential of the exponential map, expressed in suitable basis. Its computation is addressed in section \ref{sec:TheJacobian}.
	
	Note that the density with respect to $vol$ given by 
	\begin{equation}
		\label{eq:notProb}
		f(q) = h(u), 
	\end{equation}
	
	when $q=\exp_u$ and $f(q)=0$ when $q\notin \exp_p(U)$, can also be turned into a probability density by adding a global normalization factor:
	
	\begin{equation}
		\label{eq:pseudoWrapped}
		f(q) = \frac{1}{\alpha} h(u), \quad  \text{ with } \alpha = \int_{q\in \exp_p(U)}  h(\log_p(q))\mathrm{d}vol.
	\end{equation}
	

	
	Equations \eqref{eq:wrappedDensity}, \eqref{eq:notProb}, and \eqref{eq:pseudoWrapped} have sometimes been confused in the literature, see for instance \cite{srivastava2005statistical,turaga2011statistical,slama2014grassmannian,slama2015accurate}. Before focusing on wrapped densities, it is interesting to note that after being normalised, the density of \eqref{eq:pseudoWrapped} enjoys interesting properties in specific contexts. For instance, when $\mathcal{M}$ is a non compact Riemannian symmetric space of dimension $d$, the densities 
	
	\begin{equation}
		\label{RG}
		f(q\mid p,\Sigma)\propto h(\log_{p}(q))= e^{-\frac{1}{2}\log_{p}( q)^T\Sigma^{-1}\log_{p}(q)},\quad p,q\in \mathcal{M},\Sigma \in \operatornamewithlimits{SPD}(d),
	\end{equation}
	where $\log_{p}(q)$ is a coordinate expression of the inverse of the exponential map, have two remarkable properties: (i) $f$ is the maximum entropy distribution for fixed Frechet average and covariance, see \cite{pennec2006intrinsic}, and (ii) when $\Sigma$ is isotropic the maximum likelihood estimator of $p$ is the empirical Frechet average, see \cite{said2017riemannian,said2017gaussian}. However, probability densities obtained from \eqref{eq:pseudoWrapped} often suffer from several practical limitations. 1: The normalization constant can be computed explicitly only in exceptional cases. 2: Sampling from the distribution is not straightforward, and may require numerical approximations. 3: The link between the parameter $\Sigma$ and the covariance of the distribution is not explicit. 
	
	As we will see, these practical limitations do not hold for exponential wrapped densities on symmetric spaces, which makes them particularly adapted to many practical situations.. 
	
	
	

	\paragraph{Densities are explicit}
	
	On an arbitrary Riemannian manifold, such densities are hard to compute, since the exponential map and its inverse have no explicit forms. However, as we will see in the next section, on ALSS exponential maps are locally identified with Lie groups exponentials and are hence efficiently computed.
	Furthermore, we show in section \ref{sec:TheJacobian} that on these spaces, the volume distortion induced by the exponential map is always tractable. Hence, the density itself is tractable.
	
	\paragraph{Sampling is straightforward}
	
	In order to sample from $f$, it suffices to sample from $h$: if $U_1,...,U_n$ are i.i.d. random variables on a tangent space $T_p\mathcal{M}$ following the density $h$, then $X_1=\exp_p(U_1),\ldots,X_n=\exp_p(U_n)$ are i.i.d. random variables on $\mathcal{M}$ following the density $f$. Since the exponential map can be computed in closed form, exponential-wrapped densities on ALSS are trivial to sample from as long as one can sample from the pull-back density on the tangent plane.  This is in sharp contrast to the very substantial problems that are often faced in sampling from distributions supported on manifolds.
	
	\paragraph{Correspondence between moments of $f$ and $h$}
	
	A mean, or exponential barycenter, of a probability density $f$ on $\mathcal{M}$ can be defined as a point $\bar p$ satisfying
	\begin{equation*}
		\mathbb{E}_{f}[\log_{\bar p}(q)] = \int_{q\in\mathcal{M}} \log_{\bar p}(q)f(q)dvol = 0\in T_{\bar p}\mathcal{M},
	\end{equation*}
	
	see \cite{pennec2019curvature}. Hence, if the mean $\int uh(u)du$ of $h$ is $0\in T_p\mathcal{M}$, then it can be checked that $p$ is a mean of $f$. Higher intrinsic moments of the density $f$ at $p$ are usually defined as 
	\begin{equation}
		m_{p}^{k} = \mathbb{E}_{f}[\log_{p}(q)^{\otimes k}]= \int_{y\in\mathcal{M}} \log_{p}(q)^{\otimes k} f(q)dvol = \int_{u\in T_p\mathcal{M}} u^{\otimes k} h(u)du,
	\end{equation}
	where the second equality is obtained by the change of variable $u =  \log_p(q)$.
	Hence the higher moments of $f$ at $p$ are the same as those of $h$. An important consequence is that the moments of $h$ can be estimated by the empirical moments of $f$. This property does not hold for densities defined from \eqref{eq:pseudoWrapped} due to the absence of the volume correction.

	\section{Symmetric spaces}
	\label{sec:SymmetricSpaces}
	
	\subsection{Affine connections and affine locally symmetric spaces}
	
	Let $\mathcal{M}$ be a manifold endowed with an affine connection $\nabla$. Recall that the connection enables differentiation of vector fields: given two vector fields $X$ and $Y$ on $\mathcal{M}$, $(\nabla_Y X)(p)\in T_p\mathcal{M}$ defines the derivative of the field $X$ in the direction of the field $Y$ at $p$, a tangent vector at $p$.  This connection enables transportation of a vector $u\in T_{c(0)}\mathcal{M}$ along a differentiable curve $c(t)$ by imposing $\nabla_{c'(t)} u(t)=0$: this is parallel transport of $u$ along the curve $c$. A path $\gamma$ is called geodesic if $\gamma'(t)$ is the parallel transport of $\gamma'(0)$ along $\gamma$:
	$$ \nabla_{\gamma'(t)} \gamma'(t) = 0.$$
	Assume that $\gamma(0)=p$. The geodesics define an exponential map from tangent spaces to the manifold: $\exp_p(\gamma'(0))=\gamma(1)$. 
	
	Each affine connection has a torsion tensor $T$ defined as
	
	$$T(u,v) = \nabla_u v -\nabla_v u - [u,v] = 0,$$
	
	where $u,v$ are vector fields and $[.,.]$ the Lie bracket between vector fields. For every Riemannian manifold there is an affine connection which has the same geodesics and exponential maps. If the affine connection is chosen with null torsion, the connection is unique and called the Levi-Civita connection. In the rest of the paper, it is always assumed that the torsion $T$ of $\nabla$ is null:
	$$T=0.$$
	
	Though the expression of the torsion tensor does not appear explicitly in the rest of the paper, this assumption plays an important role in our main result through the equation of Jacobi fields.
	
	Affine connections also have a curvature tensor defined by
	$$ R(u,v)w  = \nabla_u \nabla_v w - \nabla_v \nabla_u w - \nabla_{[u,v]}w$$
	where $u,v,w$ are three vector fields. ALSS are defined as manifolds with an affine connection such that the derivative of the curvature tensor with respect to any vector field is always null:
	$$ \nabla R = 0.$$
	
	The assumptions $T=\nabla R=0$ encompass a large variety of spaces. An important case that we will address in the paper is when the connection $\nabla$ arise from a (pseudo-)Riemannian metric. The manifold is then called a (pseudo-)Riemannian ALSS. As is described in \cite{pennec2020beyond}, ALSS also contain another important class of spaces: arbitrary Lie groups endowed with their $0$-connection. 
	
	They are a particularly interesting class of spaces since exponentials and logarithms can be identified with matrix counterparts, and the Jacobian of the exponential can be computed explicitly.
	
	\subsection{Homogeneous symmetric spaces}
	\label{sec:AffHomSymSpaces}
	Alternatively, a homogeneous symmetric space can be characterised algebraically. It is a homogeneous space $G/K$ with an involution $\sigma$ which has the following properties: $G$ is a connected Lie group, $\sigma$ is an involutive automorphism, and $K$ is an open subgroup of the set of fixed points of $\sigma$. Such a homogeneous space has a unique canonical connection which verifies: $\nabla$ is equivariant under the action of $G$, $T=0$ and $\nabla R=0$. Hence, homogeneous symmetric spaces are also ALSS. The Lie algebra of the Lie group $G$ can be decomposed into a direct sum $\mathfrak{g} = \mathfrak{k} \oplus \mathfrak{m}$ where $\mathfrak{k}$ and $\mathfrak{m}$ are the $+1$ and $-1$ eigenspaces of $d\sigma$. Hence, $\mathfrak{k}$ is the Lie algebra of $K$, and $\mathfrak{m}$ can be identified with the tangent space at $eK$ of the quotient manifold, $\mathfrak{m}\sim T_{eK}G/K$, where $e$ is the identity of the group. 
	
	A key feature for practical use of homogeneous symmetric spaces is that for $u\in \mathfrak{m}$,  $\exp(u)K= \exp_{eK}(u)$ where the first exponential is the group exponential while the second is the exponential of the canonical affine connection, see \cite{nomizu1954invariant} section $10$. The exponential of the connection at an arbitrary point $gK$ can be computed from the Lie group exponential by
	$$\exp_{gK}(u) = g\exp(g^{-1}.u)K$$
	where $g^{-1}.u \in T_{eK}G/K\sim \mathfrak{m}$ is the action of $g^{-1}$ on the tangent vector $u\in T_{gK}G/K$.
	Another important feature is that the action of $\exp(u\in \mathfrak{m})$ on a tangent vector $v\in T_{gK}G/K$ is the parallel transport of $v$ from $gK$ to $\exp(u)gK$ along $\exp(tu)gK$.
	
	\subsection{Identifications and notations}
	\label{sec:IdentificationsAndNotations}
	
	K. Nomizu showed in \cite{nomizu1954invariant} showed that for an affine locally symmetric space $\mathcal{M}$, there is a neighborhood $N_p$ around each $p\in \mathcal{M}$ such that $N_p$ is isomorphic to a neighborhood of a homogeneous symmetric space. In the rest of the paper, 
	\begin{itemize}
		\item $\mathcal{M}$ is a differentiable manifold with a connection $\nabla$ such that $T=0$ and $\nabla R=0$, and $p$ is an arbitrary reference point
		\item $N_p$ is a neighborhood of $p$ identified to a neighborhood of a homogeneous symmetric space $G/K$ with $p\sim eK$. The tangent space $T_p\mathcal{M}$ is identified to $\mathfrak{m}$ where $\mathfrak{m}$ is defined above in section \ref{sec:AffHomSymSpaces}.
	\end{itemize}

	\section{The Jacobian of the exponential map}
	\label{sec:TheJacobian}
	
	\subsection{Main ingredient}
	\label{sec:MainIngredient}
	
	In this section we provide a general expression for the Jacobian determinant of the exponential map on ALSS. This expression is not entirely original, since it can be derived from \cite{taniguchi1984note}. However, it was never mentioned in the statistics and data science literature. Theorem \ref{thm:mainTheorem} is expressed for an arbitrary point $q\in \mathcal{M}$. We have that $\mathrm{d} \exp_q(u): T_q\mathcal{M}\rightarrow T_{\exp_q(u)}\mathcal{M}$.
	On an arbitrary ALSS, there is no reference basis, scalar product or volume measure in the tangent spaces. In order to define the Jacobian determinant 
	$$J_q(u) = \det(\mathrm{d}\exp_q(u)),$$
	
	we set an arbitrary basis $e_1,\ldots,e_d$ of $T_q\mathcal{M}$ and parallel transport it to $T_{\exp_q(u)}\mathcal{M}$ along the geodesic $\exp_q(tu)$. Check that $J_q$ is independent of the choice of basis $e_1,\ldots,e_d$ of $T_q\mathcal{M}$. Note $\tau_t:T_{q}\mathcal{M}\rightarrow T_{\exp_q(tu)}\mathcal{M}$ the parallel transport between $T_{q}\mathcal{M}$ and $ T_{\exp_q(tu)}$ along $\exp_q(tu)$.  By definition the matrix of $\tau_1$ in $e_1,..,e_d$ and $\tau_1(e_1),\ldots,\tau_1(e_d)$ is the identity, hence 
	
	$$J_q(u) = \det( \tau_1 \circ \tau_1^{-1} \circ \mathrm{d}\exp_q(u))=\det( \tau_1^{-1} \circ \mathrm{d}\exp_q(u)).$$
	Since $\tau_1^{-1} \circ \mathrm{d}\exp_q(u)$ is an endomorphism of $T_q\mathcal{M}$, its determinant is independent of a basis, hence $J_q(u)$ is independent of the basis of $T_q\mathcal{M}$.
	
	Let $R_u:T_q\mathcal{M} \rightarrow T_q\mathcal{M}$ be the linear map given by $R_u(v)=R(v,u)u$ where $R$ is the curvature tensor. Using the equation of Jacobi fields on ALSS the author of \cite{taniguchi1984note} shows that the differential of the exponential is given by:
	$$\mathrm{d}\exp_q(u) = \tau_1 \circ \sum_0^{\infty}\frac{(-R_u)^n}{(2n+1)!}.$$ 
	Triangularizing the matrix of $R_u$ over $\mathbb{C}$ leads to the following result.

	\begin{theorem}
		\label{thm:mainTheorem}
		Let $R_u$ be the linear map defined above. Note its $i$-th complex eigenvalue $\lambda_i(R_u)$ and its algebraic multiplicity $n_i$. 
		The Jacobian determinant $J_q$ of the exponential map at $u$ in the basis $e_1,..,e_d$ and $\tau_1(e_1),\ldots,\tau_1(e_d)$ is given by
		\begin{equation}
			\label{eq:JacobianMain}
			J_q(u) = \prod_i \left( \frac{\sinh\left(\sqrt{-\lambda_i(R_u)}\right)}{\sqrt{-\lambda_i(R_u)})} \right)^{n_i},    
		\end{equation}
		with $\frac{\sinh\left(\sqrt{-\lambda_i(R_u)}\right)}{\sqrt{-\lambda_i(R_u)}}=1$ when $\lambda_i(R_u)=0$. 
	\end{theorem}
	The proof is provided in appendix \ref{ap:ProofOfTheorem}. Recall that $\sinh(\mathbf i x)=\mathbf  i\sin(x)$, hence the hyperbolic sine becomes a sine when the eigenvalue $\lambda_i$ is real positive. 
	
	The formula for the case of Riemannian symmetric spaces, that can be found in \cite{helgason1979differential} page 294, has a similar structure but the eigenvalues are roots of the complexified Lie algebra $\mathfrak{g}_{\mathbb{C}}$ of $G$. Since our formula derives directly from the equation of the Jacobi fields, it is naturally expressed using the curvature tensor. A benefit is that it can be used and understood without knowledge of roots systems of semisimple Lie algebras. Nontheless, it is sometimes interesting to relate the $\lambda_i$ to algebraic quantities. The curvature tensor at the point $p$ relates to the Lie bracket of the Lie algebra of the group $G$ in a simple way, see \cite{nomizu1954invariant}:
	\begin{equation}
		\label{eq:CurveLie}
		\forall u,v,w \in T_p\mathcal{M} \sim \mathfrak{m}, \quad R(u,v,w)=-[[u,v],w].
	\end{equation}
	
	Recall also that $ad_u(v)=[u,v]$. Hence at the point $p\sim eK$, $R_u(v)=-[u,[u,v]]=-ad_u^2(v)$ and the eigenvalues of $R_u$ are the eigenvalues of $-ad_u^2$ restricted to $\mathfrak{m}$. Due to the homogeneity of $G/K$, the Jacobian $J_p$ determines the Jacobian of all other exponential maps $\exp_q$.
	\begin{corollary}
		\label{cor:pointChange}
		The Jacobian determinant of $\exp_q$ at $v$ in parallel transported basis is
		$$ J_q(v) = J_p(kg^{-1}.v),$$
		where $k$ is arbitrary element of $K$. Here $kg^{-1}.v$ is understood as the differential of the action of $kg^{-1}$ applied to $v$.
	\end{corollary}
	The proof is given in appendix \ref{ap:ProofOfTheorem}. This formula enables one to always turn the computation of the Jacobian into a computation of eigenvalues of $-ad^2$. In the rest of the paper the Jacobian $J_p$ is simply noted $J$.
	
	The formula Eq.(\ref{eq:JacobianMain}) is given in parallel transported basis and does not rely on other properties of the connection $\nabla$ other than $T=0$ and $\nabla R=0$. 
	In sections \ref{sec:Riemannian} and \ref{sec:Liegroups} 
	we give particular attention to two classes of symmetric spaces: (pseudo-)Riemannian locally symmetric spaces and Lie groups endowed with their Cartan-Schouten connection. In both contexts the additional structures enable one to state adapted results for the construction of exponential-wrapped probability densities. We address the use of the Jacobian for exponential wrapped densities on arbitrary locally symmetric spaces in section \ref{sec:GeneralSpace}.

	\subsection{Riemannian and pseudo Riemannian symmetric spaces}
	\label{sec:Riemannian}
	
	Assume that the connection $\nabla$ of the manifold $\mathcal{M}$ is the Levi-Civita connection of a Riemannian or pseudo Riemannian metric $g$. $\mathcal{M}$ has a natural volume measure $vol$ induced by the metric. Let $e_1,..,e_d\in T_p\mathcal{M}$ be an orthonormal basis ($|g(e_i,e_j)|=\delta_{ij}$), and let $\nu_p$ denote the corresponding Lebesgue measure. Since parallel transport is an isometry, the Jacobian determinant $J$ is related to the volume change of Eq.\ref{eq:wrappedDensity} in the following way,
	$$ \frac{\mathrm{d} \exp_{p*}(\nu_p)}{\mathrm{d} vol}(\exp(u))= |J(u)|^{-1}.$$
	
	We now provide the expression of the Jacobian on an example of a Riemannian symmetric spaces: real Grassmanian manifolds, and an example of a pseudo Riemannian symmetric space: pseudo-hyperboloids. We are currently not aware of references containing these formulas. Moreover, the Jacobian on Grassmannians was omitted at several occasions in the densities of wrapped distributions, see for instance \cite{srivastava2005statistical,turaga2011statistical,slama2014grassmannian,slama2015accurate}.   
	
	\subsubsection{Real Grassmanians}
	\label{subsubsec:Real Grassmanians}
	\paragraph{The Grassmanian of vector subspaces}

	The Grassmanian $Gr_k(n)$ denotes the spaces of $k$ dimensional vector subspaces of $\mathbb{R}^n$. We first describe the homogeneous symmetric structure of Grassmanians, as done in section \ref{sec:AffHomSymSpaces} for the general case. 
	
	Let $O$ denote the groups of orthogonal matrices and $SO$ their subgroups of determinant $1$. Clearly $O(n)$ acts transitively on subspaces of dimension $k$. Furthermore it is easy to see that block diagonal matrices with the first block in $O(k)$ and the second in $O(n-k)$ leave stable the vector spaces spanned by the $k$ first basis vectors. Hence $$Gr_k(n)\sim O(n)/(O(k)\times O(n-k)).$$
	This quotient can be simplified to $Gr_k(n)\sim SO(n)/S(O(k)\times O(n-k))$, where $S(O(k)\times O(n-k))$ refers to the block diagonal matrices of determinant $1$, with the first block in $O(k)$ and the second in $O(n-k)$. 
	
	The involutive automorphism of the symmetric structure is given by
	\begin{equation}
		\label{eq:involutionGrass}
		\sigma(X) = \begin{pmatrix}I_k&0\\0^T&-I_{n-k}\\ \end{pmatrix} X \begin{pmatrix}I_k&0\\0^T&-I_{n-k}\\ \end{pmatrix}.    
	\end{equation}
	It can be checked that $S(O(k)\times O(n-k))$ is an open subgroup of the set of fixed points of $\sigma$, hence the involution makes $SO(n)/S(O(k)\times O(n-k))$ a homogeneous symmetric space. Since $S(O(k)\times O(n-k))$ is compact, the quotient admits an invariant Riemannian metric and is a homogeneous Riemannian symmetric space, see \cite{helgason1979differential}. 
	
	The Lie algebra of $SO(n)$ is decomposed on eigenspaces of $\mathrm{d}\sigma$ at identity,
	\begin{equation*}
		\mathfrak{so}(n)=\mathfrak{k} \oplus \mathfrak{m}.
	\end{equation*}
	It can be checked that the $-1$ eigenspace $\mathfrak{m}$ is given by,
	\begin{equation*}
		\mathfrak{m}=\left\{ X_B= \begin{pmatrix}0&B\\-B^T&0\\ \end{pmatrix},\quad B\in Mat_{k,n-k}(\mathbb{R})   \right\}
	\end{equation*}
	where $Mat_{k,n-k}(\mathbb{R})$ are real $k$ by $n-k$ matrices. Recall that on matrix groups $ad_X(Y)=XY-YX$. The computations shown in appendix \ref{ap:Grassmann} of the eigenvalues of the adjoints $ad^2$ restricted to $\mathfrak{m}$ lead to the following Jacobian at $p\sim I_n.S(O(k)\times O(n-k))$,
	\begin{equation}
		\label{eq:JacobianGrassmann}
		J(X_B)=\prod_{i<j} \frac{\sin(\sigma_i-\sigma_j)}{\sigma_i-\sigma_j}\frac{\sin(\sigma_i+\sigma_j)}{\sigma_i+\sigma_j}\prod_i \left(\frac{\sin(\sigma_i)}{\sigma_i}\right)^{|n-2k|},
	\end{equation}
	
	where $\sigma_i$ are the singular values of $B$ counted with multiplicity one, and where each fraction is replaced by $1$ when the denominator is $0$. 
	
	Note that $S(O(k)\times O(n-k))$ has two components and that the identity component $SO(k)\times SO(n-k)$ is also an open subgroup of the set of fixed points of $\sigma$. Hence $SO(n)/(SO(k)\times SO(n-k))$ is another homogeneous symmetric space: the oriented real Grassmanian. Since $\mathfrak{m}$ and $ad_{X_B}$ remain the same, the Jacobian also has the same expression.
	
	\paragraph{The Grassmanian of affine subspaces} Let $\operatorname{Graff}_k(n)$ be the set of affine subspaces of dimension $k$ of $\mathbb{R}^n$. It is clear that the set of isometries of $\mathbb{R}^n$, noted $E(n)$, acts transitively on $\operatorname{Graff}_k(n)$. Furthermore, the stabilizer of the subspace generated by the first $k$ vectors is given by $E(k) \times O(n-k)$: a rigid motion of the subspace and a rotation of the complement. Hence, $\operatorname{Graff}_k(n)$ is a homogeneous space,
	$$\operatorname{Graff}_k(n) \sim E(n)/(E(k)\times O(n-k)).$$
	
	Authors of \cite{lim2021grassmannian} show that the geometry of this quotient is nicely described by an embedding in the Grassmanian of vector subspaces $\operatorname{Gr}_{k+1}(n+1)$. Let $V$ be a $k$ dimensional vector subspace and $b$ be a vector of $\mathbb{R}^n$. The following map $j$:
	$$j\left((V,b)\right)= \operatorname{span}(V\cup \{b,e_{n+1}\}),  $$
	where $e_{n+1}$ is the last basis vector of $\mathbb{R}^{n+1}$, embeds $\operatorname{Graff}_k(n)$ in $\operatorname{Gr}_{k+1}(n+1)$. The canonical Riemannian metric on $\operatorname{Graff}_k(n)$ is then the metric induced by $\operatorname{Gr}_{k+1}(n+1)$. Furthermore, $j(\operatorname{Graff}_k(n))$ is an open subset of $\operatorname{Gr}_{k+1}(n+1)$. Since $\operatorname{Gr}_{k+1}(n+1)$ is homogeneous Riemannian symmetric, this embedding makes $\operatorname{Graff}_k(n)$ a Riemannian locally symmetric space. Hence we have locally $\operatorname{Graff}_k(n)\sim \operatorname{Gr}_{k+1}(n+1) \sim O(n+1)/(O(k+1)\times O(n-k))$ and the Jacobian can be computed with Eq.\ref{eq:JacobianGrassmann}.
	
	\subsubsection{Pseudo-hyperboloids}
	
	We now provide the Jacobian on pseudo-hyperboloids.\\ They are pseudo-Riemannian manifolds recently used in \cite{law2020ultrahyperbolic}, where the authors show their relevance for graph embedding problems. Let us start by describing pseudo-hyperboloids, following the approach of \cite{law2020ultrahyperbolic}.
	
	For $p,q\geq 0$, let $\mathbb{R}^{p,q+1}$ be the space $\mathbb{R}^{p}\times \mathbb{R}^{q+1}$ endowed with the pseudo-Euclidean scalar product 
	$$ \langle x,y \rangle = \sum_{i=1}^{p} x_iy_i - \sum_{j=p+1}^{p+q+1} x_iy_i.$$
	Define the pseudo-hyperboloid $\mathcal{Q}_{\beta}^{p,q}$ as 
	$$ \mathcal{Q}_{\beta}^{p,q}= \{ x\in \mathbb{R}^{p,q+1} , \langle x,x \rangle = \beta \},$$
	where $\beta<0$. Pseudo-spheres are defined with $\beta>0$, but note that $\mathcal{Q}_{\beta}^{p,q}$ and $\mathcal{Q}_{-\beta}^{q+1,p-1}$ are anti-isometric. Furthermore, since all $\beta<0$ lead to homotetic pseudo-hyperboloids, we set $\beta=-1$. 
	
	As described in section \ref{sec:AffHomSymSpaces},
	we can now exhibit the symmetric space structure of $\mathcal{Q}_{\beta}^{p,q}$ and compute the Jacobian determinant of the exponential map. Let $O(p,q+1)$ be the indefinite orthogonal group which preserves the pseudo scalar product of $\mathbb{R}^{p,q+1}$. The group $O(p,q+1)$ acts transitively by isometries on $\mathcal{Q}_{-1}^{p,q}$. Since  the stabiliser of the last basis vector  $e_{p+q+1}\in \mathcal{Q}_{-1}^{p,q}$ is  the subgroup $O(p,q)$, we have that $\mathcal{Q}_{-1}^{p,q} \sim O(p,q+1)/O(p,q)$. Consider an involution similar to the one defined in Eq.\ref{eq:involutionGrass}:
	\begin{equation}
		\label{eq:involutionHyp}
		\sigma(X) = \begin{pmatrix}-I_{p+q}&0\\0&1\\ \end{pmatrix} X \begin{pmatrix}-I_{p+q}&0\\0&1\\ \end{pmatrix}.
	\end{equation}
	$\sigma$ is an involution  of $O(p,q+1)$ and it can be checked that $O(p,q)$ is an open subgroup of the set of fixed points of $\sigma$. Hence it gives $O(p,q+1)/O(p,q)$ a homogeneous symmetric structure. The Lie algebra can be decomposed on the eigenspaces of $\mathrm{d}\sigma$ at identity,
	\begin{equation*}
		\mathfrak{o}(p,q+1)=\mathfrak{k} \oplus \mathfrak{m},
	\end{equation*}
	and it can be checked that the $-1$ eigenspace $\mathfrak{m}$ is given by,
	\begin{equation*}
		\mathfrak{m}=\left\{ X_{v,w} =\begin{pmatrix}0&0&v\\0&0&w\\v^T&-w^T&0 \end{pmatrix},\quad v\in \mathbb{R}^p,w\in \mathbb{R}^q   \right\}.
	\end{equation*}
	
	Again, on a matrix group $ad_X(Y)=XY-YX$. The computations of the eigenvalues of the $ad^2$ restricted to $\mathfrak{m}$ given in appendix \ref{ap:PseudoHyp} lead to the following Jacobian at $p\sim I_{p+q+1}.O(p,q)$,
	\begin{equation}
		\label{eq:JacobianHyp}
		J(X_{v,w})=\left\{ \begin{matrix}  \left( \frac{\sinh( \|v\|^2 - \|w\|^2 )}{\|v\|^2 - \|w\|^2} \right)^{p+q-1} & \text{ if } \|v\|\neq \|w \| \\ 1  & \text{ if } \|v\|= \|w \| \end{matrix} \right.,
	\end{equation}
	where $\|v\|$ and $\|w\|$ are the Euclidean norms of $v$ and $w$.
	
	\subsection{Lie groups}
	\label{sec:Liegroups}
	
	As pointed out by \cite{taniguchi1984note}, the differential of the exponential map on symmetric spaces can be used to derive the differential of the exponential map on Lie groups. We describe here how the Jacobian determinants relate to each other. 
	
	Remarkably, every Lie group has an affine connection $\nabla$ compatible with the group structures, called the $0$-connection, which makes it an ALSS. Since the symmetric structure of the $0$-connection was only described very recently in the data science literature, see \cite{pennec2020beyond}, we recall the most important facts.  Let $\mathcal{X}$ be a manifold equipped with a Lie group structure with identity $o\in \mathcal{X}$.
	
	\begin{proposition}
		\label{prop:LieProposition}
		Let $\nabla$ be the bi-invariant connection defined by 
		$$ \nabla_{\tilde u} \tilde v = \frac{1}{2} [u,v], \quad u,v\in T_o\mathcal{X}$$
		where $\tilde u$ and $\tilde v$ are the left invariant vector fields generated by $u$ and $v$, and $[.,.]$ is the Lie bracket associated with the Lie group structure on $\mathcal{X}$. $\nabla$ is called the $0$-connection, or $0$-Cartan-Schouten connection. We have,
		\begin{itemize}
			\item[i)] $(\mathcal{X},\nabla)$ is an affine locally symmetric space
			\item[ii)] one parameter subgroups are geodesics: at $o$, the group exponential and the exponential of the connection coincide
			\item[iii)] the curvature and the Lie bracket are related by 
			$$\forall u,v,w\in T_o\mathcal{X} \quad  R(u,v,w) = -\frac{1}{4}[[u,v],w]$$
			\item[iii)] the parallel transport from $o$ to $\exp(u)$ of the vector $v$ is given by
			$$ \Pi_o^{\exp(u)}v = \mathrm{d}L_{\exp(\frac{u}{2})}\circ\mathrm{d}R_{\exp(\frac{u}{2})} (v), $$
			where $\mathrm{d}L$ and $\mathrm{d}R$ are the differential of the left and right multiplications.
		\end{itemize}
	\end{proposition}
	Proofs can be found in \cite{pennec2020beyond}. Assume now that the manifold $\mathcal{M}$ has a Lie group structure of identity $p$, and that $\nabla$ is the $0$-Cartan-Shouten connection. We show how Theorem \ref{thm:mainTheorem} leads to the formula of the Jacobian on Lie groups, given for instance in \cite{falorsi2019reparametrizing}. Note that the Lie bracket $[.,.]$ in Proposition \ref{prop:LieProposition} is not the same as the one coming from the identification $T_p\mathcal{M}\sim \mathfrak{m} \subset T_eG$, where $G$ is the Lie group involved in the local identification $\mathcal{M}\sim G/H$. Similarly to Eq.\ref{eq:CurveLie}, using iii) we can write $R_u(v)=-\frac{1}{4}[u,[u,v]]=-\frac{1}{4}ad_u^2(v)$, but where $ad_u$ is now the adjoint map of the Lie algebra of $\mathcal{M}$. This relation enables us to obtain an algebraic expression of the Jacobian which involves only the structure of $\mathcal{M}$ and not of the Lie group $G$. 
	
	On a Lie group the differential of the group exponential is usually computed in basis transported by left (or right) multiplication. As a result, the Jacobian determinant is a volume change between a Lebesgue measure on the Lie algebra and a Haar measure. As $iii)$ shows, parallel transported basis are not simply obtained by left or right multiplication, hence the Jacobian determinant of Theorem \ref{thm:mainTheorem} is not a volume change with respect to a Haar measure. Though, a simple calculation shown in appendix \ref{ap:LieGroups} enables us to relate the two Jacobians and to obtain the following corollary of Theorem \ref{thm:mainTheorem}.

	\begin{corollary}
		\label{cor:LieGroups}
		Set a basis $B=(e_1,...,e_d=u)$ of $T_p\mathcal{M}$ and let $u\in T_p\mathcal{M}$. Let $B_L$ be the basis of $T_{\exp(u)}\mathcal{M}$ obtained by left multiplication of $B$. The Jacobian determinant of the exponential map expressed in $B$ and $B_L$ is given by  
		
		$$ \tilde J_p(u) = \prod_i \left( \frac{1-e^{\alpha_i(u)}}{\alpha_i(u)} \right)^{m_i},$$
		where $\alpha_i(u)$ are the eigenvalues of $ad_u$ and $m_i$ their multiplicities.
	\end{corollary}
	
	The proof is given in appendix \ref{ap:LieGroups}. Let $\nu_p$ be the Lebesgue measure on $T_p\mathcal{M}$ and $vol$ be the left Haar measure on $\mathcal{M}$ generated by the basis $B$. We have
	$$\frac{\mathrm{d} \exp_*(\nu_p)}{\mathrm{d} vol}(\exp(u))= |\tilde J_p(u)|^{-1}.$$
	
	Similarly to corollary \ref{cor:pointChange} expressing the Jacobian at arbitrary points, it can be checked that on Lie groups, the Jacobian in left-transported basis computed at an arbitrary $q$ is given by
	$$\tilde J_q(u) = \tilde J_p(\mathrm{d}L_{q^{-1}}.u).$$

	\subsection{The general case}
	\label{sec:GeneralSpace}
	
	We are currently not aware of practical problems in data science or physics involving a random phenomenon on a symmetric space which is not Riemannian, pseudo Riemannian, or a Lie group.
	However, such spaces remain an interesting class, with some interesting special cases.
	For example, the connection $\nabla$ on $\mathbb{R}^2$ whose Christoffel's coefficients are all zeros except $\Gamma^2_{11}(x,y)=y$ is symmetric but does not correspond to a Riemannian or Lie structure. We outline the use of the Jacobian on a general affine locally symmetric space but we do not provide proofs of the results in this paper.
	
\begin{figure}[h]
		\label{Fig:holonomy}
		\begin{center}
			\includegraphics[scale=0.2]{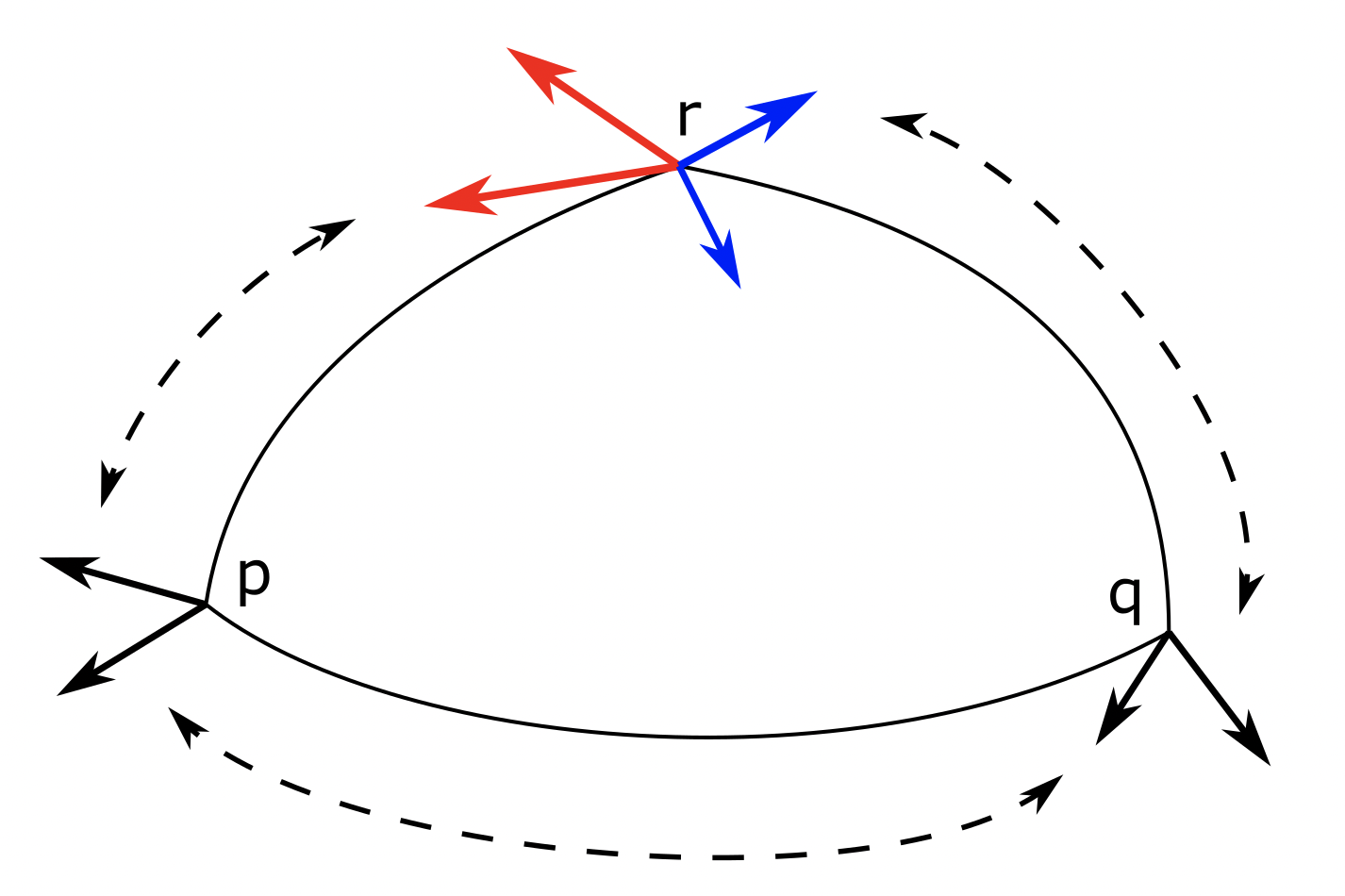}
			\caption{The dotted lines represent the parallel transport between $p$, $q$ and $r$. The term $H$ in Eq.\ref{eq:generalCase} is the determinant of the red basis in the blue basis.}
		\end{center}
	\end{figure}

	In both the Riemannian and Lie group settings we interpreted the Jacobian as a volume change between a Lebesgue measure of the tangent space and a reference measure on the manifold. On general symmetric spaces there might not be such a reference measure. In that case,  exponential wrapped probability distributions do not have a natural notion of density, even when they are absolutely continuous with respect to the Lebesgue measures of the charts of $\mathcal{M}$. However, relative densities between exponential wrapped probability distributions can still be computed. Let $p,q\in \mathcal{M}$ and $\mathcal{U}\subset \mathcal{M}$ be such that $\log_p$ and $\log_q$ are well defined on $\mathcal{U}$. Let $\lambda_p$ and $\lambda_q$ be two probability distributions supported on $\log_p(\mathcal{U})$ and $\log_q(\mathcal{U})$ respectively. Set an arbitrary reference basis on $T_p\mathcal{M}$ and parallel transport it to $T_q\mathcal{M}$. If $\lambda_p$ and $\lambda_q$ have densities $h_p$ and $h_q$ with respect to the corresponding Lebesgue measures, then for any $r\in \mathcal{U}$ it can be shown that 
	
	\begin{equation}
		\label{eq:generalCase}
		\frac{\mathrm{d}\exp_{q*}\lambda_q}{\mathrm{d}\exp_{p*}\lambda_p}(r)=H. \frac{J_p(\log(r))}{J_q(\log_p(r))} \frac{h_q(\log_q(r))}{h_p(\log_p(r))},
	\end{equation}
	$$ $$
	
	where $H$ is the determinant of the holonomy map along the geodesic triangle $r\rightarrow q \rightarrow p\rightarrow r$, see Fig.\ref{Fig:holonomy}. Furthermore as mentioned in section \ref{sec:AffHomSymSpaces}, the parallel transport on a symmetric space is obtained by the action of elements of $\exp(\mathfrak{m})$. This enables us to compute $H$ explicitly.

	\section{A classification experiment using exponential wrapped distributions}
	\label{sec:A classification experiment using exponential wrapped distributions}

	
	Outside cases where laws are modeled using a fixed tangent space, the analysis of the convergence of density estimators based on exponential-wrapped distributions is still in early development. 
	
	As suggested in section \ref{sec:wrappedDistributions}, exponential-wrapped distributions are sometimes conveniently estimated with moment matching estimators. The study of the theoretical properties of moment matching estimators is out of the scope of this paper. However, the module\\ \textbf{frechet\_mean} of the python package \textbf{Geomstats}, see \cite{miolane2020geomstats}, now enables the computation of the empirical moments on several symmetric spaces in a simple way. 
	
	Taking advantage of this python package, we present a classification experiment on simulated data drawn in two Riemannian symmetric spaces: the real Grassmannian of two-dimensional subspaces of $\mathbb{R}^4$ and the space of $2\times 2$ symmetric positive definite matrices. 
	
	On both spaces, we consider four equiprobable classes. For a class $C_i$ a training set and a test set are drawn from an exponential-wrapped density $f_i$. Each training set is then modeled by an estimated exponential-wrapped distribution $\hat f_i$, and samples from test sets are classified according the maximum a posteriori probability. Several approaches are compared, depending on the number, and location, of tangent spaces used to model the data. In \textit{model 1}, the training sets are modeled with exponential wrapped distributions originating from different tangent spaces, while in \textit{model 2} and \textit{3}, the training sets are modeled with exponential wrapped distributions originating from the same tangent space. In \textit{model 2} and \textit{3}, the Jacobians between the tangent spaces and the manifold are not involved in the classification, since all the data are classified in the same tangent space.  The classification results show the interest of \textit{model 1} over \textit{model 2} and \textit{model 3}. All the computations necessary to the classification are achieved with the package \textbf{Geomstats}.

	The training set and test set of the class $C_i$ are obtained by sampling from an isotropic exponential-wrapped normal density $f_i =f(.;p_i,v_i) $, which we describe in the next paragraph.
	
	\subsection{Isotropic exponential-wrapped normal distributions}
	\label{subsec:Isotropic exponential-wrapped normal distributions}
	
	Define the distribution\\ $\mathcal{N}_{\mathcal{M}}(p,v)$ as
	$$ \mathcal{N}_{\mathcal{M}}(p,v) = \exp_{p*}\left( \mathcal{N}\left(0,\frac{v}{d} \langle.,.\rangle_{p}\right) \right), $$
	
	where $\mathcal{N}$ is a multivariate normal distribution, $\langle.,.\rangle_{p}$ is the inner product of $T_{p}\mathcal{M}$, and $d$ the dimension of $\mathcal{M}$. Note $f(.;p,v)$ the density of $\mathcal{N}_{\mathcal{M}}(p,v)$. When the manifold $\mathcal{M}$ is a space of symmetric positive definite matrices, the exponential map is a bijection between each tangent spaces and $\mathcal{M}$. After particularizing Eq. \ref{eq:wrappedDensity}, we obtain that the density $f_i(.) = f(.;p_i,v_i)$ is given by
	\begin{equation}
		\label{eq:tangentNormal}
		f_i(q) = \frac{1}{J_{p_i}(\log_{p_i}(q))}\frac{1}{\sqrt{(2\pi w_i)^d}} e^{-\frac{d(q,p_i)^2}{2w_i}},    
	\end{equation}
	
	where $w_i = \frac{v_i}{d}$. When $\mathcal{M}$ is a real Grassmannian manifold, the exponential maps are surjective but not injective. In the current experiment, the normal distributions on the Grassmannian are taken with small variances, which enables to neglect the mass outside the injectivity radius. This hypothesis is often made in practice, see \cite{falorsi2019reparametrizing,mallasto2019wrapped,fletcher2003gaussian}, and avoid the technicalities of truncated normal distributions used in \cite{turaga2011statistical,slama2015accurate}.  This assumption enables to approximate the density by Eq. \ref{eq:tangentNormal}.

	As pointed out in the end of section \ref{sec:wrappedDistributions}, an important aspect of such exponential-wrapped normal density, with respect to other types of normal densities on manifolds, is that the parameters $p$ and $v$ correspond to empirical moments of $\mathcal{N}_{\mathcal{M}}(p, v)$. Indeed, the change of variable $u =  \log_p(q)$ lead to
	
	$$ \int_{\mathcal{M}} \log_p(q) f(q;p,v) \mathrm{d}vol = \frac{1}{\sqrt{(2\pi w)^d}} \int_{T_p\mathcal{M}} u e^{-\frac{\|u\|^2}{2w^2}}\mathrm{d}u = 0,$$
	
	where $vol$ is the Riemannian volume and $w=\frac{v}{d}$. Hence $p$ is a mean of 
	$\mathcal{N}_{\mathcal{M}}(p, v)$. 
	
	The same change of variable also gives
	
	$$ \int_{\mathcal{M}} d(q,p)^2 f(q;p,v) \mathrm{d}vol = \frac{1}{\sqrt{(2\pi w)^d}} \int_{T_p\mathcal{M}} \|u\|^2 e^{-\frac{\|u\|^2}{2w^2}}\mathrm{d}u = v,$$
	Hence $v$ is the variance of
	$\mathcal{N}_{\mathcal{M}}(p, v)$. This allows to estimate the parameters $p$ and $v$ by empirical moments. Note that on symmetric spaces with positive curvature, such as Grassmannian manifolds, the uniqueness of the mean is not guaranteed when the distribution is not sufficiently concentrated. Hence the convergence of the estimation of $p$ by an empirical mean is also not guarenteed. The small variance hypothesis enables to neglect this phenomenon.
	
	\subsection{The Grassmannian $Gr_2(4)$}
	
	We now give the expression of the Jacobian on the Grassmannian of two-dimensional vector subspaces of $\mathbb{R}^4$, noted $Gr_2(4)$, as well as the parameters of the four classes $C_i$. $Gr_2(4)$ is a four dimensional manifold described in section \ref{subsubsec:Real Grassmanians}. It is identified with the quotient
	
	$$ O(4) / (O(2)\times O(2)),$$
	
	and its tangent space at $I.O(2)\times O(2)$ is identified with
	
	\begin{equation*}
		\mathfrak{m}=\left\{ X_B= \begin{pmatrix}0&B\\-B^T&0\\ \end{pmatrix},\quad B\in Mat_{2,2}(\mathbb{R})   \right\}.
	\end{equation*}
	
	The Jacobian becomes
	
	$$ J(X_B)=\frac{\sin(\sigma_1-\sigma_2)}{\sigma_1-\sigma_2} \frac{\sin(\sigma_1+\sigma_2)}{\sigma_1 + \sigma_2},$$
	
	where $\sigma_1$ and $\sigma_2$ are the singular values of $B$. On $Gr_2(4)$ the parameters of the distributions of the four classes are chosen as $p_i = \exp_I(X_{B_i})$, with
	
	$$ B_1 = \begin{pmatrix} 0& 0 \\ 0& 0\end{pmatrix},B_2 = \begin{pmatrix} 0& 0 \\ -\frac{\pi}{2}& 0\end{pmatrix}, B_3 = \begin{pmatrix} 0& -\frac{\pi}{2} \\ 0& 0\end{pmatrix}, B_4 = \begin{pmatrix} 0& 0 \\ 0& -\frac{\pi}{2}\end{pmatrix},$$
	
	and 
	
	$$ v_1 = v_2 = v_3 = v_4 = 0.6. $$
	
	For this choice of variance, a Monte-Carlo sampling shows that in the tangent spaces, $99.8\%$ of the mass lies in the injectivity ball $B(\frac{\pi}{2})$ and $60\%$ lie in the ball $B(\frac{\pi}{4})$, $\frac{\pi}{2}$ being the injectivity radius of $Gr_2(4)$. This distribution of mass is consistent with the approximation made in Eq.\ref{eq:tangentNormal}, and ensures in practice the uniqueness of the mean.
	
	\subsection{The space of $2\times 2$ symmetric positive definite matrices}
	Before providing the expression of the Jacobian and the parameters of the classes $C_i$, start by a brief description of the structure of symmetric space. Note $\operatorname{Sym}(2)$ and $\operatorname{SPD}(2)$ the spaces of $2\times 2$ symmetric and symmetric positive definite matrices. Since $\operatorname{SPD}(2)$ is an open subset of the vector space $\operatorname{Sym}(2)$, all the tangent spaces of $\operatorname{SPD}(2)$ are identified with $\operatorname{Sym}(2)$. Endow $\operatorname{SPD}(2)$ with the following Riemannian metric
	
	$$ g_{\Sigma}(X,Y)=\operatorname{trace}(\Sigma^{-1}X\Sigma^{-1}Y)$$
	
	where $\Sigma \in \operatorname{SPD}(2)$ and  $X,Y \in \operatorname{Sym}(2)$. The metric $g$ makes $\operatorname{SPD}(2)$ a Riemannian symmetric space, whose detailed presentation can be found in \cite{terras1984harmonic}. Let us simply give the identifications introduced in section \ref{sec:IdentificationsAndNotations}. $\operatorname{SPD}(2)$ is identified with $GL(2)/O(2)$ by the map $\Sigma \mapsto \Sigma^{1/2}.O(2)$, where $\Sigma^{1/2}$ is the symmetric square root of $\Sigma$, and the tangent space of $GL(2)/O(2)$ at $I.O(2)$ is itself identified with 
	
	$$ \mathfrak{m}=\operatorname{Sym}(2).$$
	
	This lead to an identification of $T_I \operatorname{SPD}(2)$ and $\mathfrak{m}$ given by $X\mapsto \frac{1}{2}X$. For $X$ in $T_I \operatorname{SPD}(2)$, the computation of the eigenvalues of $ad_{\frac{1}{2}X}^2:\mathfrak{m}\rightarrow \mathfrak{m}$ give the following Jacobian,
	
	$$J(X) = 2\frac{\sinh(\frac{\sigma_1-\sigma_2}{2})}{\sigma_1-\sigma_2},$$
	
	where $\sigma_1$ and $\sigma_2$ are the eigenvalues of $X$, see also \cite{chevallier2017kernel}. On $\operatorname{SPD(2)}$, the parameters of the distributions of the four classes are chosen as
	
	$$p_1 = \exp_I\left( \begin{pmatrix} 3& 0 \\ 0& -3 \end{pmatrix} \right),  p_2 = \exp_I\left( \begin{pmatrix} 3& 0.3 \\ 0.3& -3 \end{pmatrix} \right),$$
	$$p_3 = \exp_I\left( \begin{pmatrix} -3& 0 \\ 0& 3 \end{pmatrix} \right),  p_4 = \exp_I\left( \begin{pmatrix} -3& -0.3 \\ -0.3& 3 \end{pmatrix} \right),$$
	
	and 
	$$ v_1 = v_2 = v_3 = v_4 = 2.$$

	\subsection{Estimation of exponential-wrapped normal distributions}
	
	The test sets are modeled according to three procedures. 
	
	\hfill 
	
	\begin{itemize}
		\item In \textit{model 1}, the parameters of the density $f_i = f(.;p_i,v_i)$ of the class $C_i$ are estimated by the empirical mean and variance $\hat p_i,\hat v_i$. The test set of the class $C_i$ is then modeled by the density $\hat f_i = f(.;\hat p_i,\hat v_i)$.
		\item In \textit{model 2}, all the data points are first lifted in a single tangent space $T_{p_0}\mathcal{M}$ by the logarithm $\log_{p_0}$. The point $p_0$ is chosen to be a mean of all the training sets. Each lifted training set is then modeled by an isotorpic normal density $h(.;\hat \mu_i,\hat v_i)$ on $T_{p_0}\mathcal{M}$ of parameters
		$$ \hat \mu_i = \frac{1}{N} \sum_i \log_{p_0}(q_i) \text{ and } \hat v_i = \frac{1}{N} \sum_i \|\log_{p_0}(q_i)-\mu_i\|_{p_0}^2,$$
		where $N$ is the size of the training sets.
		\item \textit{model 3} differs from \textit{model 2} in the choice of the lifting point $p_0$, which is now set as the mean of the training set of the first class. 
	\end{itemize}

	\subsection{Classification results}
	
	Data are classified according to the maximum a posteriori probability. Since we consider equiprobable classes, maximizing the posterior probability is equivalent to maximizing the likelihood of the observation. Hence, a data point at $q\in \mathcal{M}$ is classified as
	
	$$ C(q) = \begin{cases}
		argmax_i \ f(q;\hat p_i , v_i ) & \textit{(model 1)} \\
		argmax_i \ h(q;\hat \mu_i,\hat v_i ) & \textit{(model 2 and \textit{3})}.
	\end{cases}.$$
	
	On both spaces, we consider four equiprobable classes with a training set of size $4\times 50$ and a test set of size $4\times 50$. The classification is repeated $5000$ times. The following table shows the average rate of good classifications, plus or minus a standard deviation. 
	
	\hfill
	
	
	
	
	
	\begin{center}
		\begin{tabular}{|c|c|c|c|}
			\hline
			& \textit{model 1} & \textit{model 2} & \textit{model 3}\\ \hline
			
			$Gr_2(4)$ & $0.838 \pm 0.026$  & $0.777 \pm 0.031$&$0.501\pm 0.052$ \\  \hline     
			
			$\operatorname{SPD}(2)$ & $0.816 \pm 0.028$ & $0.682 \pm 0.050$ & $0.618\pm 0.035$\\ \hline
			
		\end{tabular}
	\end{center}
	\hfill
	
	On both spaces, the results illustrates the advantage of working with multiple tangent spaces over a global linearization of the space. In the case of a global linearization, choosing an off-centered tangent space (\textit{model 3}) led to lower classification results than those obtained with a centered tangent space (\textit{model 2}).

	\section{Discussion}
	\label{sec:Discussion}
	
	Exponential-wrapped distributions had previously been defined and used on specific manifolds. In this paper we showed that ALSS are a broad class of manifolds where exponential-wrapped densities can be computed in closed form, under an injectivity condition. These distributions have then been used in a classification experiment on simulated data. Further studies should investigate deeper the impact the various factors affecting the classification results, such as the curvature tensor of the manifold or the number and locations of classes. In order to provide a theoretical background to these results, future works will also focus on the study of the convergence of estimators based on exponential-wrapped distributions. An important problem remains open in the case where the tangent space used to model data is not fixed in advance: differentiating the likelihood of densities with respect to the base point of the tangent space. The differentiation involves the double exponential expansion, whose expression on arbitrary affine manifolds can be found in \cite{pennec2019curvature} section 3.2 and \cite{gavrilov2007double}. Our future efforts will focus on understanding the implications of this formula for density estimation with exponential wrapped densities. 
	
	
	\section*{Acknowledgement}
	E.C. would like to thank Salem Said, Xavier Pennec, Nicolas Guigui and Yann Thanwerdas for fruitful discussions on symmetric spaces. The authors acknowledge support for this research from an Office of Naval Research grant N00014-14-1-0245/N00014-16-1-2147.  Y.L. is supported by the US National Science Foundation under award DMS-2107934.

	\section{Appendix A : proofs of the general forms of Jacobians}
	
	\subsection{Proof of theorem \ref{thm:mainTheorem} }
	\label{ap:ProofOfTheorem}

	The main part of the proof is similar to \citep{taniguchi1984note}. Chose a basis $e_1,\ldots,e_d$ of $T_q\mathcal{M}$ with $e_d=u$.  Expressed in $e_1(0),..,e_d(0)$ and $e_1(t),..,e_d(t)$, we have
	$$ J(tu) =\det( \mathrm{d}\exp_p(tu) ) =  \det\left( \frac{\partial \exp }{\partial e_1}(te_d),...,\frac{\partial \exp }{\partial e_d}(te_d) \right).$$
	
	On manifolds with null torsion, the  
	$$ Y_i(t) = t\frac{\partial \exp }{\partial e_i}(te_d),\quad i=1,..,d.$$
	are solutions of the Jacobi equations
	$ Y_i''(t) + R_{e_d(t)}\left(Y_i(t)\right) = 0, $ 
	with initial conditions
	$Y_i(0)=0,\quad \text{and } Y_i'(0)=e_i.$
	$ Y''(t)$ refers here to the seconde covariante derivative along the geodesic $\exp(tu)$ and $R_{e_d(t)}$ to the map $R(e_d(t),.)e_d(t)$. Given a tensor field $T$ along $\exp(tu)$ note $[T]$ its coordinates in the basis $e_1(t),..,e_d(t)$. Since the manifold is locally symmetric, $[R_{e_d(t)}]=[R_{e_d(0)}]=[R_{u}]$. The Jacobi equation becomes a second order differential equation in $\RR^d$ with constant coefficients.
	
	$$ [Y_i]'' + [R_u][Y_i]=0.$$
	In the rest of the proof, the matrix $[R_u]$ is simply noted $R$. Let $F_t:\RR^d\rightarrow \RR^d$ be the linear map defined by $F_t(X)=v(t)$ with $v$ the unique solution of the Cauchy problem
	$$
	\left\{
	\begin{array}{l}
		v''+Rv=0,\\
		v(0)=0,\\
		v'(0)=X.
	\end{array}
	\right.
	$$
	
	It can be checked that $\frac{1}{t^{d}}F_t$ is the matrix expression of linear map $\mathrm{d}\exp_p(tu)$, hence $J(u)=\det (F_1)$. Turn first the differential equation into a first order differential equation. We get
	$$ \begin{pmatrix}
		v'\\v''
	\end{pmatrix} = \begin{pmatrix}
		0&I\\-R&0
	\end{pmatrix} \begin{pmatrix}
		v\\v'
	\end{pmatrix}.$$
	Let $A=\begin{pmatrix}
		0&I\\R&0
	\end{pmatrix}$, 
	the solution is given by $\begin{pmatrix}
		v(t)\\v'(t)
	\end{pmatrix}=e^{tA}\begin{pmatrix}
		v(0)\\v'(0)
	\end{pmatrix}$.
	It is easy to check that $e^{tA}=\begin{pmatrix}
		E_t&F_t\\G_t&H_t
	\end{pmatrix}$ where $F_t$ is the linear map defined previously. Hence, we want to compute the determinant of the upper right block of $e^{A}$. Compute first the powers of $A$. It can be checked by induction that for $k\in \mathbb{N}$,
	
	$$A^{2k}= t^{2k}\begin{pmatrix}
		(-R)^k&0\\0&(-R)^k
	\end{pmatrix} \text{ and } A^{2k+1}= A^{2k+1}\begin{pmatrix}
		0&(-R)^{k}\\(-R)^{k+1}&0
	\end{pmatrix}.$$
	
	We can deduce that 
	$F_1 = 0 + I +0+ \frac{-R}{3!} + 0 + \frac{(-R)^2}{5!}+..+0+ \frac{(-R)^{2k}}{(2k+1)!}+...$, which is analogous to the formula provided in \citep{taniguchi1984note}. Hence we have that the matrix of the differential of the exponential map at the tangent vector $u$ in parallel transported basis is given by 
	$$\mathrm{d}\exp_p(u) =  \sum_0^{\infty}\frac{(-R)^n}{(2n+1)!}.$$ 
	
	Recall that a matrix can always be triangularized over $\mathbb{C}$. Let $R=PTP^{-1}$ with $T$ an upper triangular matrix. Recall also that the diagonal elements of $T$ are the complex eigenvalues $\lambda_i$ of $R$. We have $\mathrm{d}\exp_p(u) = P \left( \sum_0^{\infty}\frac{(-T)^n}{(2n+1)!}\right) P^{-1}$, and $\det(\mathrm{d}\exp_p(u))=\det\left( \sum_0^{\infty}\frac{(-T)^n}{(2n+1)!}\right)$. Since $ \sum_0^{\infty}\frac{(-T)^n}{(2n+1)!}$ is upper triangular we have that
	
	\begin{equation}
		\label{eq:detF}
		J(u)=\det(\mathrm{d}\exp_p(u))= \prod_i  \left(\sum_0^{\infty}\frac{(-\lambda_i)^n}{(2n+1)!}\right)^{n_i}=\prod_i \left(\frac{ \sinh( \sqrt{-\lambda_i})}{ \sqrt{-\lambda_i}}\right)^{n_i},
	\end{equation}
	where $n_i$ is the multiplicity of $\lambda_i$, and where $\frac{\sinh(0)}{0}=1$. Note that since $\sinh$ is odd, it is clear that changing the choice of square root $\sqrt{-\lambda_i}$ to $-\sqrt{-\lambda_i}$ does not affect the determinant.

	\subsection{The Jacobian at arbitrary base-point : proof of corollary \ref{cor:pointChange} }
	
	Recall that the connection $\nabla$ is equivariant under the action of $G$. Hence $g.\exp_q(u)=\exp_{g.q}(g.u)$ and $g.\mathrm{d}\exp_q(u)=\mathrm{d}\exp_{g.q}(g.u)$. Let $B$ be a basis of $T_{p}\mathcal{M}$ and $B'$ its parallel translation to $T_{\exp_p(u)}$ along $\exp_p(u)$. We have that $g.B$ and $g.B'$ are basis of $T_{g.q}\mathcal{M}$ and $T_{g.\exp_p(g.u)}\mathcal{M}$ and that the determinant of $\mathrm{d}\exp_q(u)$ in $B$ and $B'$ is the same as the determinant of $g.\mathrm{d}\exp_q(u)$ in $g.B$ and $g.B'$. Moreover, the equivariance of the connection gives that $g.B$ and $g.B'$ are also  related by parallel transport. In other words,
	$ J_q(u) = J_{g.q}(g.u)$.
	In section \ref{sec:MainIngredient}, the function $J$ is defined at $p\sim eK$. Hence if $q\sim gK$, we have $kg^{-1}.gK=eK$ for all $k\in K$ and
	
	$$ J_q(u) = J(kg^{-1}.u). $$

	\subsection{The Lie group formula : proof of corollary \ref{cor:LieGroups}}
	\label{ap:LieGroups}
	
	At $\exp(u)$, the two basis $B_L$ and $B_T$ obtained by respectively by left invariance and parallel transport are given by: $B_L = \mathrm{d}L_{\exp(u)}.B$ and $B_T=\mathrm{d}L_{\exp(\frac{u}{2})}\mathrm{d}R_{\exp(\frac{u}{2})}B$.
	We have $\tilde J(u) =  J(u).\det_{B_L}(B_T)$, hence we need to compute $\det_{B_L}(B_T)$:
	$$ \det_{B_L}(B_T) = \det_{\mathrm{d}L_{\exp(-u)}B_L}(\mathrm{d}L_{\exp(-u)}B_T)=\det_{e_1,..,e_d}(\mathrm{d}L_{\exp(-\frac{u}{2})}\mathrm{d}R_{\exp(\frac{u}{2})}.B)=\det(Ad_{\exp(-\frac{u}{2})}). $$
	Since $Ad_{\exp(-\frac{u}{2})}=e^{-\frac{1}{2}ad_u}$, $\det_{B_L}(B_T) = e^{-\frac{1}{2}\sum_i \alpha_i(u)}$. On the other hand, 
	$$ e^{- \frac{1}{2} \alpha_i(u)}  \frac{\sinh\left(\frac{1}{2}\alpha_i(u)\right)}{\frac{1}{2}\alpha_i(u)}
	=e^{- \frac{1}{2} \alpha_i(u)} 2 \frac{e^{\frac{1}{2}\alpha_i(u)}-e^{-\frac{1}{2}\alpha_i(u)}}{2\alpha_i(u)}  
	=\frac{1-e^{-\alpha_i(u)}}{\alpha_i(u)},$$
	which lead to the desired formula.

	\section{Appendix B : Eigenvalues of $ad_X^2$ on specific examples}
	
	\subsection{The real Grassmannian}
	\label{ap:Grassmann}
	
	Let $X_{B_1} =\begin{pmatrix}0&B_1\\-B_1^T&0 \end{pmatrix}$ and $X_{B_2} =\begin{pmatrix}0&B_2\\-B_2^T&0 \end{pmatrix}$. Let us first compute $ad_{X_{B_1}}(X_{B_2})$. We obtain
	$$ad_{X_{B_1}}(X_{B_2}) = X_{B_1}X_{B_2}-X_{B_2}X_{B_1}\\
	=\begin{pmatrix} B_2B_1^T-B_1B_2^T&0\\ 0& B_2^TB_1-B_1^TB_2 \end{pmatrix}. $$
	
	Hence,
	\begin{eqnarray*}
		ad_{X_{B_1}}(ad_{X_{B_1}}(X_{B_2})) &=&\begin{pmatrix}0& 2B_1B_2^TB_1-B_1B_1^TB_2-B_2B_1^TB_1 \\B_1^TB_1B_2^T+B_2^TB_1B_1^T-2B_1^TB_2B_1^T&0 \end{pmatrix}.
	\end{eqnarray*}
	
	Let $\varphi_{B_1}(B_2)=2B_1B_2^TB_1-B_1B_1^TB_2-B_2B_1^TB_1 $. Let $B_1 = U D V$ be the singular value decomposition of $B_1$. We have
	\begin{eqnarray*}
		\varphi_{B_1}(B_2) &=&2B_1B_2^TB_1-B_1B_1^TB_2-B_2B_1^TB_1\\
		&=&   2U D VB_2^TU D V-U D VV^{-1}D^TU^{-1}B_2-B_2V^{-1}D^TU^{-1}U D V,
	\end{eqnarray*}
	\begin{eqnarray*}
		\varphi_{B_1}(UB_2V) &=&2U D VV^{-1}B_2^TU^{-1}U D V-U DD^TU^{-1}UB_2V-UB_2VV^{-1}D^TD V\\
		&=&2U D B_2^T D V-U DD^TB_2V-UB_2D^TD V\\
		&=&U(2 D B_2^T D - DD^TB_2-B_2D^TD )V = U\varphi_{D}(B_2)V,
	\end{eqnarray*}
	
	which shows that the eigenvalues of $\varphi_{B_1}$ and $\varphi_{D}$ are the same. We assume now $B_1=D$ with diagonal $\sigma_1,...,\sigma_q$, where $q=\min(k,n-k)$. Let $E_{i,j}$ be the canonical basis of $k$ by $n-k$ matrices. Assume $i,j\leq q$, a short calculation shows that
	\begin{eqnarray*}
		\varphi_{D} (E_{ii}) &=&0\\
		\varphi_{D} (E_{ij}+E_{ji})&=&-(\sigma_i-\sigma_j)^2(E_{ij}+E_{ji})\\
		\varphi_{D} (E_{ij}-E_{ji})&=&-(\sigma_i+\sigma_j)^2(E_{ij}-E_{ji}).
	\end{eqnarray*}
	
	When $k>n-k$, we can have $i>n-k$, and
	$\varphi_{D} (E_{ij})=-\sigma_j^2E_{ij}$,
	while when $k<n-k$ and $j>k$,
	$\varphi_{D} (E_{ij})=-\sigma_i^2E_{ij}$.
	Hence the singular value $\sigma_i$ appears $k-(n-k)$ times or $(n-k)-k$ times. Since $R_u(v)=-\operatorname{ad}_u^2(v)$, Eq.\ref{eq:JacobianMain} can be rewritten with  $\sqrt{-\lambda_i(R_u)}=\sqrt{\lambda_i(\operatorname{ad}_u^2)}$ and the Jacobian becomes
	\begin{equation}
		J(X_B)=\prod_{i<j} \frac{\sin(\sigma_i-\sigma_j)}{\sigma_i-\sigma_j} \frac{\sin(\sigma_i+\sigma_j)}{\sigma_i+\sigma_j}\prod_i \left(\frac{\sin(\sigma_i)}{\sigma_i}\right)^{|n-2k|}.
	\end{equation}
	
	\subsection{Pseudo-hyperboloids}
	\label{ap:PseudoHyp}
	
	Let $X_{v_1,w_1} =\begin{pmatrix}0&0&v_1\\0&0&w_1\\v_1^T&-w_1^T&0 \end{pmatrix}$ and\\ 
	$X_{v_2,w_2} =\begin{pmatrix}0&0&v_2\\0&0&w_2\\v_2^T&-w_2^T&0 \end{pmatrix}$. Let us first compute $ad_{X_{v_1,w_1}}(X_{v_2,w_2})$. We obtain
	$$ad_{X_{v_1,w_1}}(X_{v_2,w_2}) = X_{v_1,w_1}X_{v_2,w_2}-X_{v_2,w_2}X_{v_1,w_1}\\
	=\begin{pmatrix}v_1v_2^T-v_2v_1^T&v_2w_1^T-v_1w_2^T&0\\w_1v_2^T-w_2v_1^T&w_2w_1^T-w_1w_2^T&0\\0&0&0 \end{pmatrix}. $$
	
	
	Hence we have $ad_{X_{v_1,w_1}}(ad_{X_{v_1,w_1}}(X_{v_2,w_2})) =...$
	$$ \begin{pmatrix}0&0&0 \\0&0&0\\v_1^Tv_1v_2^T-v_1^Tv_2v_1^T- w_1^Tw_1v_2^T+w_1^Tw_2v_1^T&v_1^Tv_2w_1^T-v_1^Tv_1w_2^T  -w_1^Tw_2w_1^T+w_1^Tw_1w_2^T&0 \end{pmatrix}+$$
	
	$$ \begin{pmatrix}0&0&-v_1v_2^Tv_1+v_2v_1^Tv_1 -v_2w_1^Tw_1+v_1w_2^Tw_1 \\0&0& -w_1v_2^Tv_1+w_2v_1^Tv_1 - w_2w_1^Tw_1+w_1w_2^Tw_1\\0&0&0 \end{pmatrix}.$$
	Note $\varphi$ the map on vectors $\begin{pmatrix}v_2\\w_2 \end{pmatrix}\in \mathbb{R}^{p+q}$ induced by $ad_{X_{v_1,w_1}}^2$. The matrix of $\varphi$ is 
	$$ M_{\varphi}= \begin{pmatrix} -v_1v_1^T+(\|v_1\|^2 - \|w_1\|^2)I_p &v_1w_1^T\\-w_1v_1^T&w_1w_1^T+(\|v_1\|^2 - \|w_1\|^2)I_q\end{pmatrix}$$
	
	Hence, $M_{\varphi}=\begin{pmatrix}v_1\\w_1 \end{pmatrix}\begin{pmatrix}-v_1\\w_1 \end{pmatrix}^T +(\|v_1\|^2 - \|w_1\|^2)I$. The matrix $A=\begin{pmatrix}v_1\\w_1 \end{pmatrix}\begin{pmatrix}-v_1\\w_1 \end{pmatrix}^T$ is rank one and has $0$ as eigenvalue with multiplicity at least $p+q-1$. When $(\|w_1\|^2 - \|v_1\|^2)\neq 0$, the matrix $A$ can be diagonalized with $(\|w_1\|^2 - \|v_1\|^2)$ in the first index and $0$ on the rest of the diagonal. $M_{\varphi}$ can then be diagonalized with $0$ in the first index and $(\|v_1\|^2 - \|w_1\|^2)$ on the $p+q-1$ remaining indices. When $(\|w_1\|^2 - \|v_1\|^2)=0$, the matrix $A$ is not diagonalizable by only has eigenvalue $0$. Hence $M_{\varphi}$ only has eigenvalue $0$. Since the $0$ eigenvalues do not affect the Jacobian, it can always be written as
	$$
	J(X_{v,w})=  \left( \frac{\sinh( \|v\|^2 - \|w\|^2 )}{\|v\|^2 - \|w\|^2} \right)^{p+q-1}.$$

	\bibliographystyle{apalike}
	\bibliography{ref.bib}

\begin{thebibliography}{}

\bibitem[Chevallier et~al., 2015]{chevallier2015probability}
Chevallier, E., Barbaresco, F., and Angulo, J. (2015).
\newblock Probability density estimation on the hyperbolic space applied to
  radar processing.
\newblock In {\em International Conference on Geometric Science of
  Information}, pages 753--761. Springer.

\bibitem[Chevallier et~al., 2016]{chevallier2016kernel}
Chevallier, E., Forget, T., Barbaresco, F., and Angulo, J. (2016).
\newblock Kernel density estimation on the {S}iegel space with an application
  to radar processing.
\newblock {\em Entropy}, 18(11):396.

\bibitem[Chevallier et~al., 2017]{chevallier2017kernel}
Chevallier, E., Kalunga, E., and Angulo, J. (2017).
\newblock Kernel density estimation on spaces of {G}aussian distributions and
  symmetric positive definite matrices.
\newblock {\em SIAM Journal on Imaging Sciences}, 10(1):191--215.

\bibitem[Ding and Regev, 2020]{ding2020deep}
Ding, J. and Regev, A. (2020).
\newblock Deep generative model embedding of single-cell rna-seq profiles on
  hyperspheres and hyperbolic spaces.
\newblock {\em Nature communications}, 12(1):1--17.

\bibitem[Falorsi et~al., 2019]{falorsi2019reparametrizing}
Falorsi, L., de~Haan, P., Davidson, T.~R., and Forré, P. (2019).
\newblock Reparameterizing distributions on {L}ie groups.
\newblock In {\em 22nd International Conference on Artificial Intelligence and
  Statistics (AISTATS), PMLR: Volume 89}, pages 3244--3253.

\bibitem[Fisher, 1953]{fisher1953dispertion}
Fisher, R.~A. (1953).
\newblock Dispersion on a sphere.
\newblock {\em Proceedings of the Royal Society of London. Series A.
  Mathematical and Physical Sciences}, 217(1130):295--305.

\bibitem[Fletcher et~al., 2003]{fletcher2003gaussian}
Fletcher, P.~T., Joshi, S., Lu, C., and Pizer, S. (2003).
\newblock Gaussian distributions on lie groups and their application to
  statistical shape analysis.
\newblock In {\em Biennial International Conference on Information Processing
  in Medical Imaging}, pages 450--462. Springer, Berlin, Heidelberg.

\bibitem[Gavrilov, 1954]{gavrilov2007double}
Gavrilov, A.~V. (1954).
\newblock The double exponential map and covariant derivation.
\newblock {\em Siberian Mathematical Journal}, 48(1):56--61.

\bibitem[Hall et~al., 1987]{hall1987kernel}
Hall, P., Watson, G., and Cabrera, J. (1987).
\newblock Kernel density estimation with spherical data.
\newblock {\em Biometrika}, 74(4):751--762.

\bibitem[Hauberg, 2018]{hauberg2018gaussian}
Hauberg, S. (2018).
\newblock Directional statistics with the spherical normal distribution.
\newblock In {\em 21st International Conference on Information Fusion (FUSION).
  IEEE}, pages 704--711.

\bibitem[Helgason, 1979]{helgason1979differential}
Helgason, S. (1979).
\newblock {\em Differential Geometry, Lie Groups, and Symmetric Spaces}.
\newblock Academic press.

\bibitem[Hendriks, 1990]{hendriks1990nonparametric}
Hendriks, H. (1990).
\newblock Nonparametric estimation of a probability density on a {R}iemannian
  manifold using {F}ourier expansions.
\newblock {\em The Annals of Statistics}, pages 832--849.

\bibitem[Huckemann et~al., 2010]{huckemann2010mobius}
Huckemann, S.~F., Kim, P.~T., Koo, J.-Y., Munk, A., et~al. (2010).
\newblock M{\"o}bius deconvolution on the hyperbolic plane with application to
  impedance density estimation.
\newblock {\em The Annals of Statistics}, 38(4):2465--2498.

\bibitem[Jona-Lasino et~al., 2012]{jona2012spatial}
Jona-Lasino, G., Gelfand, A., and Jona-Lasino, M. (2012).
\newblock Spatial analysis of wave directional data using wrapped gaussian
  processes.
\newblock {\em The Annals of Applied Statistics}, 6(4):1478–1498.

\bibitem[Kato and McCullagh, 2020]{kato2020shogo}
Kato, S. and McCullagh, P. (2020).
\newblock Some properties of a cauchy family on the sphere derived from the
  möbius transformations.
\newblock {\em Bernoulli}, 26(4):3224--3248.

\bibitem[Kim, 1998]{kim1998deconvolution}
Kim, P.~T. (1998).
\newblock Deconvolution density estimation on {SO}({N}).
\newblock {\em The Annals of Statistics}, 26(3):1083--1102.

\bibitem[Kurtek et~al., 2012]{kurtek2012statistical}
Kurtek, S., Srivastava, A., Klassen, E., and Ding, Z. (2012).
\newblock Statistical modeling of curves using shapes and related features.
\newblock {\em Journal of the American Statistical Association}, 107,
  No.499:1152--1165.

\bibitem[Law and Stam, 2020]{law2020ultrahyperbolic}
Law, M.~T. and Stam, J. (2020).
\newblock Ultrahyperbolic representation learning.
\newblock In {\em Advances in Neural Information Processing Systems 33 (NeurIPS
  2020)}.

\bibitem[Lim et~al., 2021]{lim2021grassmannian}
Lim, L.-H., Wong, K. S.-W., and Ye, K. (2021).
\newblock The {G}rassmannian of affine subspaces.
\newblock {\em Foundations of Computational Mathematics}, 21:537–574.

\bibitem[Mallasto and Feragen, 2018]{mallasto2019wrapped}
Mallasto, A. and Feragen, A. (2018).
\newblock Wrapped {G}aussian process regression on {R}iemannian manifolds.
\newblock In {\em Proceedings of the IEEE Conference on Computer Vision and
  Pattern Recognition}, pages 5580--5588.

\bibitem[Mallasto et~al., 2019]{mallasto2019probabilistic}
Mallasto, A., Hauberg, S., and Feragen, A. (2019).
\newblock Probabilistic {R}iemannian submanifold learning with wrapped
  {G}aussian process latent variable models.
\newblock In {\em 22nd International Conference on Artificial Intelligence and
  Statistics (AISTATS), PMLR: Volume 89}, pages 2368--2377.

\bibitem[Mardia, 1972]{mardia1972statistics}
Mardia, K.~V. (1972).
\newblock {\em Statistics of {D}irectional {D}ata}.
\newblock Academic Press.

\bibitem[Mathieu et~al., 2019]{mathieu2019continuous}
Mathieu, E., Le~Lan, C., Maddison, C.~J., Tomioka, R., and Teh, Y.~W. (2019).
\newblock Continuous hierarchical representations with {P}oincar{\'e}
  variational auto-encoders.
\newblock In {\em Advances in Neural Information Processing Systems}, pages
  12544--12555.

\bibitem[Miolane et~al., 2020]{miolane2020geomstats}
Miolane, N., Guigui, N., Brigant, A.~L., Mathe, J., Hou, B., Thanwerdas, Y.,
  Heyder, S., Peltre, O., Koep, N., Zaatiti, H., Hajri, H., Cabanes, Y.,
  Gerald, T., Chauchat, P., Shewmake, C., Brooks, D., Kainz, B., Donnat, C.,
  Holmes, S., and Pennec, X. (2020).
\newblock Geomstats: A python package for riemannian geometry in machine
  learning.
\newblock {\em Journal of Machine Learning Research}, 21(223):1--9.

\bibitem[Nava-Yazdani et~al., 2020]{nava2019geodesic}
Nava-Yazdani, E., Hege, H.-C., Sullivan, T.~J., and von Tycowicz, C. (2020).
\newblock Geodesic analysis in {K}endall’s shape space with epidemiological
  applications.
\newblock {\em Journal of Mathematical Imaging and Vision}, 62:549–559.

\bibitem[Nomizu, 1954]{nomizu1954invariant}
Nomizu, K. (1954).
\newblock Invariant affine connections on homogeneous spaces.
\newblock {\em American Journal of Mathematics}, 76(1):33--65.

\bibitem[Pelletier, 2005]{pelletier2005kernel}
Pelletier, B. (2005).
\newblock Kernel density estimation on {R}iemannian manifolds.
\newblock {\em Statistics \& Probability Letters}, 73(3):297--304.

\bibitem[Pennec, 2006]{pennec2006intrinsic}
Pennec, X. (2006).
\newblock Intrinsic statistics on riemannian manifolds: Basic tools for
  geometric measurements.
\newblock {\em Journal of Mathematical Imaging and Vision}, 25(127).

\bibitem[Pennec, 2019]{pennec2019curvature}
Pennec, X. (2019).
\newblock Curvature effects on the empirical mean in {R}iemannian and affine
  manifolds: a non-asymptotic high concentration expansion in the small-sample
  regime.
\newblock {\em arXiv preprint arXiv:1906.07418}.

\bibitem[Pennec and Lorenzi, 2020]{pennec2020beyond}
Pennec, X. and Lorenzi, M. (2020).
\newblock {\em "Beyond Riemannian geometry: The affine connection setting for
  transformation groups", Riemannian Geometric Statistics in Medical Image
  Analysis}.
\newblock Science Direct.

\bibitem[Said et~al., 2017a]{said2017riemannian}
Said, S., Bombrun, L., Berthoumieu, Y., and Manton, J.~H. (2017a).
\newblock Riemannian {G}aussian distributions on the space of symmetric
  positive definite matrices.
\newblock {\em IEEE Transactions on Information Theory}, 63(4):2153--2170.

\bibitem[Said et~al., 2017b]{said2017gaussian}
Said, S., Hajri, H., Bombrun, L., and Vemuri, B.~C. (2017b).
\newblock Gaussian distributions on {R}iemannian symmetric spaces: statistical
  learning with structured covariance matrices.
\newblock {\em IEEE Transactions on Information Theory}, 64(2):752--772.

\bibitem[Slama et~al., 2014]{slama2014grassmannian}
Slama, R., Wannous, H., and Daoudi, M. (2014).
\newblock {G}rassmannian representation of motion depth for 3d human gesture
  and action recognition.
\newblock In {\em 22nd International Conference on Pattern Recognition}, pages
  3499--3504.

\bibitem[Slama et~al., 2015]{slama2015accurate}
Slama, R., Wannous, H., Daoudi, M., and Srivastava, A. (2015).
\newblock Accurate 3d action recognition using learning on the {G}rassmann
  manifold.
\newblock {\em Pattern Recognition}, 48(2):556--567.

\bibitem[Srivastava et~al., 2005]{srivastava2005statistical}
Srivastava, A., Joshi, S., Mio, W., and Liu, X. (2005).
\newblock Statistical shape analysis: Clustering, learning, and testing.
\newblock {\em Transactions on Pattern Analysis and Machine Intelligence},
  27(4):590--602.

\bibitem[Taniguchi, 1984]{taniguchi1984note}
Taniguchi, H. (1984).
\newblock A note on the differential of the exponential map and {J}acobi fields
  in a symmetric space.
\newblock {\em Tokyo Journal of Mathematics}, 7(1):177--181.

\bibitem[Terras, 1984]{terras1984harmonic}
Terras, A. (1984).
\newblock {\em Harmonic Analysis on Symmetric Spaces and Applications II}.
\newblock Springer.

\bibitem[Turaga et~al., 2011]{turaga2011statistical}
Turaga, P., Veeraraghavan, A., Srivastava, A., and Chellappa, R. (2011).
\newblock Statistical computations on {G}rassmann and {S}tiefel manifolds for
  image and video-based recognition.
\newblock {\em Transactions on Pattern Analysis and Machine Intelligence},
  33(11):2273 -- 2286.

\end{thebibliography}
	
\end{document}